\documentclass[a4paper,reqno]{article}

\usepackage{amsthm}
\usepackage{amsmath}\usepackage{amssymb}
\usepackage{enumitem}\setlist{nolistsep}
\usepackage{color}

\usepackage{a4wide}

\usepackage{hyperref}

\usepackage{graphicx}\usepackage{float}

\newcommand{\dd}{{\rm{d}}}

\newcommand{\im}{{\mathrm{i}}}
\newcommand{\Up}{{\mathcal U}_+}
\newcommand{\up}{u_+}
\newcommand{\lp}{\lambda_+}
\newcommand{\G}{{\mathcal{G}}}
\newcommand{\U}{{\mathcal{U}}}
\newcommand{\V}{{\mathcal{V}}}

\newcommand{\h}{{\mathcal{H}}}

\renewcommand{\d}{{\mathrm{d}}}

\begin{document}

\title{Cut-and-paste for impulsive gravitational waves with $\Lambda$:\\ The
geometric picture}
\author{
  Ji\v{r}\'{\i}~Podolsk\'y$^1$\thanks{{\tt podolsky@mbox.troja.mff.cuni.cz}},
  Clemens~S\"amann$^2$\thanks{{\tt clemens.saemann@univie.ac.at}},
  Roland~Steinbauer$^2$\thanks{{\tt roland.steinbauer@univie.ac.at}}
  and Robert~\v{S}varc$^1$\thanks{{\tt robert.svarc@mff.cuni.cz}} \\ \\
  $^1$ Institute of Theoretical Physics,\\
  Charles University, Faculty of Mathematics and Physics, \\
  V Hole\v{s}ovi\v{c}k\'ach 2, 18000 Prague 8, Czech Republic.\\ \\
  $^2$ Faculty of Mathematics, University of Vienna, \\
  Oskar-Morgenstern-Platz 1, 1090 Vienna, Austria. \\ \\
}

\maketitle

\begin{abstract}
Impulsive gravitational waves in Minkowski space were introduced by Roger
Penrose at the end of the 1960s, and have been widely studied  over the decades.
Here we focus on non-expanding waves which later have
been generalised to impulses travelling in all
constant-curvature backgrounds, that is also the (anti-)de Sitter universe.
While Penrose's original construction was based on his vivid geometric
`scissors-and-paste' approach in a flat background, until now a comparably
powerful visualisation and understanding have been missing in the
${\Lambda\not=0}$ case. In this work we provide such a picture: The (anti-)de
Sitter hyperboloid is cut along the null wave surface, and the `halves' are then
re-attached with a suitable shift of their null generators across the wave
surface. This special family of global null geodesics defines an appropriate
comoving coordinate system, leading to the continuous form of the metric.
Moreover, it provides a complete understanding of the nature of the Penrose
junction conditions and their specific form. These findings shed
light on recent discussions of the memory effect in impulsive waves.
\medskip

\medskip


\end{abstract}

\section{Introduction}

\begin{figure}[htb]\centering
 \includegraphics[width=.67\textwidth]{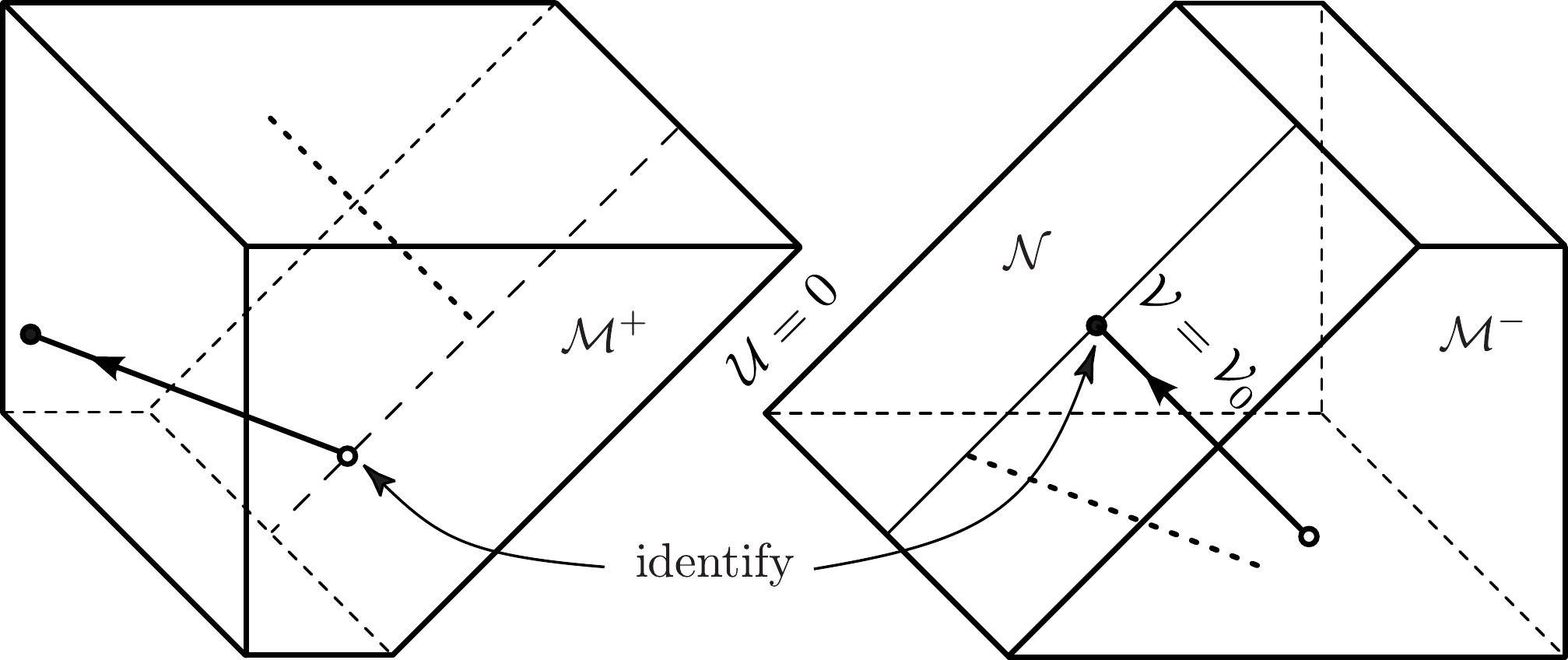}
 \caption{The construction of impulsive \textit{pp}-waves with the
cut-and-paste method: Minkowski space is cut along the null hyperplane
$\mathcal{N}$ given here by ${\U=0}$, and the two `halves' ${\mathcal M}^\pm$ are
then re-attached with a shift along the null generators of $\mathcal{N}$.
We have also depicted a null geodesic of ${\mathcal M}^-$ given by ${\V=\V_0}$, 
hitting  $\mathcal{N}$. Instead of continuing unbroken
into $\mathcal{M}^+$ (indicated by the dashed line), the interaction with the
impulse does not only make it jump due to the identification of the points but
also refracts it, see the bold line in ${\mathcal M}^+$.
}
\label{fig:planecut}
\end{figure}

Impulsive gravitational waves are theoretical models of short but violent bursts
of gravitational radiation, which have been introduced by Roger Penrose in the
late 1960s \cite[p.\ 189ff]{P:68}, \cite[p.\ 82ff]{P:68a}. Subsequently, in his
seminal 1972 paper \cite{Pen:72} he studied their geometry most explicitly using
a vivid `scissors-and-paste' approach\footnote{Nowadays more commonly referred
to as cut-and-paste method.} to construct impulsive \emph{pp}-waves:
Minkowski space is cut along a null hyperplane into two `halves',  which are
subsequently re-attached with a warp, given by the so-called Penrose junction
conditions, which guarantee that the field equations are satisfied everywhere,
see Figure \ref{fig:planecut}.

Penrose also presented these spacetimes in alternative forms using continuous or
distributional metrics, for an overview see e.g.\ \cite[Chapter 20]{GP:09},
\cite{BH:03}, \cite{P:02}. Moreover, in a classical paper Peter Aichelburg and
Roman Sexl \cite{AS:71} showed that certain impulsive \emph{pp}-waves arise by
boosting a Schwarzschild black hole to the speed of light. This AS-procedure has
subsequently been applied to other geometries of the Kerr-Newman family, leading
to more general impulses. The corresponding geometries have in turn been used as
a playground for particle scattering at the Planck scales, see  e.g.\
\cite{LS:94,S:18}.

In another landmark paper \cite{HT:93}, Masahiro Hotta and Masahiro Tanaka in
1993 have managed to include a non-vanishing cosmological constant $\Lambda$ in
the AS-procedure, thereby overcoming severe conceptional intricacies. They
boosted the Schwarzschild-(anti-)de Sitter solution to the speed of light to
obtain a non-expanding impulsive gravitational wave generated by  null monopole
particles propagating in (anti-)de Sitter space (abbreviated as (A)dS from now
on). Their key-technique was to consider the boost in a convenient
representation of (A)dS as hyperboloid in a $5$-dim flat space.

In the late 1990s Jerry Griffiths together with the first author
systematically studied the entire class of non-expanding impulsive gravitational
waves travelling in (A)dS universe \cite{PG:97,P:98,PG:99,PG:99a}, giving also
an illustrative description of the geometry of the wave surfaces. These are
spherical and hyperboloidal for ${\Lambda>0}$ and ${\Lambda<0}$, respectively,
see also \cite{P:02,PO:01}. The explicit construction of the continuous form of
the metric \cite{PG:99} was based on the use of conformally flat coordinates.
In this approach the cosmological constant does not explicitly appear in the
junction conditions, but an analogue of the vivid picture of Penrose's
cut-and-paste construction was not considered. In this work we
present such a picture: The (A)dS hyperboloid is first cut along a
null-hypersurface into two `halves'. These are then re-attached with the `dual'
null generators shifted along the generators of the wave surface by a suitable
shift, which also changes their speeds. To obtain the details of this
construction it is essential to understand the specific behaviour of the null
geodesic generators of the (A)dS hyperboloid upon crossing the impulsive wave.

Here we draw on recent works \cite{SSLP:16,SS:17} which are in turn based on the
analysis of \cite{PO:01}. The fact that it is essentially the \emph{behaviour of
the null geodesics} that \emph{determines the Penrose junction conditions}, was
first explicitly noted for ${\Lambda=0}$ in \cite{Ste:98} and put to use in
\cite{KS:99}. To emphasise this, we have included a null geodesic perpendicular
to the wave surface $\mathcal{N}$ in our Figure \ref{fig:planecut}. It is not
only shifted along the null generator of the wave surface but its speed also
changes upon its interaction with the wave, see the `jump formulas' e.g.\
\cite[Thm.\ 3]{KS:99a}. The paper \cite{KS:99} also provides a framework to
address the Penrose junction conditions and the corresponding `discontinuous
coordinate transformation' in a mathematically rigorous way. The generalisation
of the corresponding approach to the case of non-vanishing $\Lambda$ is now also
possible. Its technical implementation is, however, deferred to a separate, more
mathematically minded paper.
\medskip

This letter is organized in the following way: In the next section we summarise
the necessary prerequisites, introducing the various forms of the non-expanding
impulsive wave metrics, thereby fixing our notations. In the main technical
section \ref{sec:nullgeo} we study in detail the interaction of the null
geodesic generators of the (A)dS hyperboloid with the wave impulse. Then we
present our main result, which is the geometric account on the Penrose junction
conditions in section \ref{sec:sap}. Finally, we discuss the way in which our
analysis leads to a mathematically watertight treatment of the `discontinuous
coordinate transformation' relating the ($4$-dim) continuous and the
distributional forms of this large class of impulsive wave metrics to one
another.

\section{Non-expanding impulsive waves with $\Lambda$}

To begin, we give a brief account on the various construction methods of
non-expanding impulsive gravitational waves with $\Lambda$ following
\cite[Chapter 20]{GP:09}.
We start with the conformally flat form of the constant-curvature
(i.e., Minkowski and (A)dS) backgrounds
\begin{equation}\label{backgr}
  \d s_0^2= \frac{2\,\d\eta\,\d\bar\eta
    -2\,\d{\mathcal U}\,\d{\mathcal
      V}}{[\,1+{\frac{1}{6}}\Lambda(\eta\bar\eta-{\mathcal U}{\mathcal
      V})\,]^2}\,,
\end{equation}
where ${\mathcal U}, \mathcal{V}$ are the usual null and $\eta,\bar\eta$ the
usual complex spatial coordinates. Next we apply the coordinate transformation
separately for negative and positive values of ${\mathcal U}$,
\begin{equation}\label{ro:trsf}
  {\mathcal U}=u\,,\quad
  {\mathcal V}=
  \begin{cases}
    v &\mbox{for ${\mathcal U}<0$}\\
    v+h+u\,h_{,Z}h_{,\bar Z}  &\mbox{for ${\mathcal U}>0$}
  \end{cases}
  \,,\quad
  \eta=
  \begin{cases}
    Z &\mbox{for ${\mathcal U}<0$}\\
    Z+u\,h_{,\bar Z} &\mbox{for ${\mathcal U}>0$}
  \end{cases}\,,
\end{equation}
where $h(Z,\bar Z)$ is any real-valued function. This leads to the
\emph{continuous} form of the metric  \cite{P:98,PG:99}
\begin{equation}\label{conti}
  \dd s^2= \frac{2\,|\dd Z+\up(h_{,Z\bar Z}\dd Z+{h}_{,\bar Z\bar Z}\dd\bar
    Z)|^2-2\,\dd u\dd v}{[\,1+\frac{1}{6}\Lambda(Z\bar Z-uv+\up G)\,]^2}\,,
\end{equation}
where\footnote{This choice of sign of $G$ is in accordance with
\cite{PO:01,SSLP:16,SS:17}, which are our main points of reference, but
different from e.g.\ \cite{PSSS:15}.} ${G(Z,\bar Z)= Zh_{,Z}+\bar Zh_{,\bar
Z}-h}$ and $\up\equiv \up(u)=0$ for $u\leq 0$ and $\up(u)=u$ for $u\geq0$ is the
\emph{kink   function}. Hence the metric \eqref{conti} is of local Lipschitz
regularity in $u$ which is beyond the reach of classical smooth Lorentzian
geometry. The latter reaches down to $C^{1,1}$ --- at least as far as convexity
and causality are concerned \cite{Min:15,KSS:14,KSSV:14,GGKS:18}. However, the
Lipschitz property is decisive since it prevents the most dramatic downfalls in
causality theory which are known to occur for H\"older continuous metrics
\cite{HW:51,CG:12,SS:18,GKSS:19,Min:19}. More specifically, in the context of
the initial value problem for the geodesic equation, the Lipschitz property
guarantees the existence of solutions \cite{Ste:14} which, due to the special
geometry of the models at hand, are even (globally) unique \cite{PSSS:15}.
\medskip

The corresponding \emph{distributional} metric for non-expanding impulsive
waves in any constant-curvature background has the form \cite{PV:98}
\begin{equation}
  \dd s^2= \frac{2\,\dd\eta\,\dd\bar\eta-2\,\dd \U\,\dd
    \V+2\h(\eta,\bar\eta)\,\delta({\U})\,\dd {\U}^2}
  {[\,1+\frac{1}{6}\Lambda(\eta\bar\eta-{\U}{\V})\,]^2} \,, \label{4D_imp}
\end{equation}
where $\delta$ is the Dirac-measure and $\h(\eta,\bar\eta)$ is a real-valued
function.
Actually, it can \emph{formally} be derived from \eqref{conti} by rewriting
\eqref{ro:trsf} in the unified form of a  `discontinuous coordinate 
transformation'
using the Heaviside function $\Theta$ as
\begin{align} \label{trans}
  {\U}&=u\,,\nonumber \\
  {\V}&=v+\Theta(u)\,h+\up\,h_{,Z}h_{,\bar Z}\,, \\
  \eta&=Z+\up\,h_{,\bar Z}\,,\nonumber
\end{align}
with the specification
\begin{equation}\label{eq:H}
  h(Z,\bar Z)=\h(\eta,\bar \eta)|_{u=0}\,,
\end{equation}
and keeping all distributional terms.
Although the metric \eqref{4D_imp} possesses a distributional coefficient and
consequently lies beyond Geroch-Traschen's `maximal sensible class' of
metrics \cite{GT:87}, by their simple geometric structure the (distributional)
 curvature can be computed unambiguously leading to an impulsive
 component
$\Psi_4=(1+\frac{1}{6}\Lambda\eta\bar\eta)^2\,\h_{,\eta\eta}\,\delta(\U)$.
\medskip

The \emph{Penrose junction conditions} can now be explicitly obtained from the
transformation \eqref{trans}. More precisely, defining ${{\mathcal
    M}^-=\{\U<0\}}$ and  ${{\mathcal M}^+=\{\U>0\}}$ we can read off the
following relation by taking the limits $u=\U\nearrow 0$ and
$u=\U\searrow 0$ in \eqref{trans}
\begin{equation}
  \Big(\U=0,\V,\eta,\bar\eta\Big)_{{\mathcal
M}^-}\,=\,\Big(\U=0,\V-\h(\eta,\bar\eta),\eta,\bar\eta\Big)_{{\mathcal M}^+}\,.
\end{equation}
While the `discontinuous transformation' \eqref{trans} thus provides a clear
geometric picture, by its lack of smoothness, it is mathematically not
meaningful. Nevertheless, in the special case of
\textit{pp}-waves (for ${\Lambda=0}$), nonlinear distributional geometry
\cite[Ch.\ 3]{GKOS:01} was invoked to provide a rigorous mathematical
interpretation
of \eqref{trans} as a `generalised diffeomorphism' \cite{KS:99a}. There the key
step was a detailed study of the geodesics in the distributional form of the
metric. This study has recently been generalized in \cite{SSLP:16,SS:17} to the
$\Lambda\not=0$ case, and we will employ these results as
well as those of \cite{PO:01} to give a clear geometric interpretation of the
transformation \eqref{trans} in the general case.
\medskip

These works  all use a convenient $5$-dim
formalism, and here we will also employ it to explicitly study the null geodesic
generators of the (A)dS hyperboloid and their interaction with the impulsive
wave. Specifically, we consider the $5$-dim  impulsive \textit{pp}-wave
manifold
\begin{equation}\label{5D_imp}
  \dd s^{2}=-2\dd U \dd V+\dd Z_{2}^{2}+\dd Z_{3}^{2}+ \sigma\dd Z_{4}^{2}
  +H(Z_{2},Z_{3},Z_{4})\delta(U)\dd U^{2} \,,
\end{equation}
with the constraint
\begin{align}
  -2UV+{Z_2}^{2}+{Z_3}^{2}+ \sigma {Z_4}^{2} =\sigma a^{2}
  \label{Constraint_Hyp} \,,
\end{align}
and the parameters given by $\sigma=\pm 1 = \mathrm{sign}\, \Lambda$ and
$a=\sqrt{3 \sigma /\Lambda}$. Using the metric (\ref{5D_imp}) with
(\ref{Constraint_Hyp}) we can easily see the geometry of the
impulsive wave spacetime: The impulse is located on the null
hypersurface  ${U=0}$, which is
\begin{equation}
  {Z_2}^2+{Z_3}^2+\sigma{Z_4}^2=\sigma a^2\,,\label{imp_surface}
\end{equation}
corresponding to a non-expanding 2-sphere for ${\Lambda>0}$ and a hyperboloidal
2-surface for ${\Lambda<0}$.
The manifold \eqref{5D_imp}, \eqref{Constraint_Hyp} is related to the $4$-dim
form (\ref{4D_imp}) via
\begin{equation}
  {U} = \frac{\U}{\Omega}\,, \qquad   {V} = \frac{\V}{\Omega} \,, \qquad
  Z_2+iZ_3= \sqrt2\,\frac{\eta}{\Omega}\equiv \frac{x+i y}{\Omega}\,,
 \qquad Z_4 = a \left(\frac{2}{\Omega}-1\right), \label{CoordTrans_4D_to_5D}
\end{equation}
where we use the abbreviation
%\begin{equation}
$\Omega=1+{\textstyle\frac{1}{6}}\Lambda(\eta\bar\eta-\U\V)
\equiv 1+{\textstyle\frac{1} {12}}\Lambda(x^2+y^2-2\U\V)% \,,
$
%\end{equation}
%\quad\mbox{
and the associated real coordinates $x,y$ %with $\eta,\bar\eta$, i.e.,
\begin{equation}
  \eta=\frac{1}{\sqrt{2}}\,(x+iy)\,,
\end{equation}
as well as the relation%}\quad
\begin{equation}
H =
\frac{2\h}{1+\frac{1}{6}\Lambda\eta\bar\eta}
\equiv\frac{2\h}{1+\frac{1}{12} \Lambda(x^2+y^2)}
\,. \label{HhRelation}
\end{equation}

\section {Explicit null geodesics}\label{sec:nullgeo}
The key point is to analyse in detail the null geodesics in the impulsive wave
spacetimes at hand. Since we ultimately are interested in understanding the
transformation \eqref{trans} relating the two $4$-dim  forms \eqref{conti},
\eqref{4D_imp} of the metric to one another, we start with the geodesics of
\eqref{4D_imp} `before' the impulse, i.e., for $\U<0$ . Then we will turn to the
$5$-dim  representation \eqref{5D_imp}, \eqref{Constraint_Hyp} for two distinct
reasons: We have to base our approach on the rigorous results on such geodesics
in this representation \cite{PO:01,SSLP:16,SS:17} rather than on the analysis
using the $4$-dim  approach given in \cite{PSSS:15}, which uses the continuous
form of the metric \eqref{conti}. Indeed, in the latter the transformation
\eqref{trans} was explicitly used to \emph{derive} the form of the geodesics and
hence such an approach  would lead to a circular argument. Second, the $5$-dim
approach leads to a very vivid picture paralleling the one given in Figure
\ref{fig:planecut}.
\medskip

\subsection{Background null geodesics in the $4$-dim form (\ref{4D_imp})}
We first explicitly determine a convenient family of null geodesics of the
(A)dS background \eqref{backgr}, since away from the impulse they will coincide
with null geodesics in the impulsive wave \eqref{4D_imp}.
To be consistent with the standard ${\Lambda=0}$ case, we choose the following
family of geodesics in \emph{flat} space using the coordinates ${(\U,\V,x,y)}$,
\begin{equation}\label{simp4DGeod}
\gamma(\tilde{\lambda}) = \left(
\begin{array}{c}
\tilde{\lambda}\\
\V_0\\
x_0\\
y_0\\
\end{array}\right)\,.
\end{equation}
These are parametrised by the three real constants $\V_0$, $x_0$, and $y_0$.
They give the position of the geodesic at the parameter value
${\U=\tilde\lambda=0}$, which, in the later context, will be the instant when
the wave impulse is crossed. Next we apply the conformal transformation to
explicitly derive the corresponding null geodesics of the (A)dS background
\eqref{backgr}. To obtain an \emph{affine} parameter $\lambda$, we use the
relation
\begin{equation}
\frac{\dd \lambda}{\dd \tilde{\lambda}} = c\,\Omega^{-2} \,, \label{NewAff}
\end{equation}
with ${c=\hbox{const}}$, and
\begin{equation}
\Omega(\tilde\lambda)=1+\frac{\Lambda}{12}\left(x_0^2+y_0^2-2\tilde{\lambda}
\V_0\right) = \alpha + \beta\,\tilde{\lambda} \,,
\end{equation}
where we have defined
\begin{equation}
\alpha\equiv 1+\frac{\Lambda}{12}\left(x_0^2+y_0^2\right) \,, \qquad
\beta\equiv -\frac{\Lambda}{6}\,\V_0 \,. \label{AlphaBetaDef}
\end{equation}
To explicitly integrate (\ref{NewAff}), we distinguish two cases depending on
the value of ${\beta}$ (which is proportional to ${\V_0}$):
\begin{enumerate}[noitemsep]
\item[(1)] ${\beta=0}$: Here we trivially obtain
%\begin{equation}
$\lambda = \frac{c}{\alpha^2}\,\tilde{\lambda}+d$, i.e., % \,, \qquad
%\Leftrightarrow \qquad
$\tilde{\lambda}= \frac{\alpha^2}{c}\,(\lambda-d)$,
%\end{equation}
and it is natural to set ${d=0}$ to have ${\lambda=0}$ iff
${\tilde{\lambda}=0}$. The ansatz (\ref{simp4DGeod}) thus leads to the null
geodesics
\begin{equation}
\gamma({\lambda}) = \left(
\begin{array}{c}
\alpha^2 \lambda/c\\
\V_0\\
x_0\\
y_0\\
\end{array}\right)\,. \label{4DGeodAfBeta0}
\end{equation}

\item[(2)] ${\beta\neq0}$: In this case
%\begin{equation}
$\lambda=-\frac{c}{\beta}(\alpha+\beta\tilde{\lambda})^{-1}+d$, i.e.,
%\,, \qquad \Leftrightarrow \qquad
$\tilde{\lambda}=-\frac{c}{\beta^2(\lambda-d)}-\frac{\alpha}{\beta}$,
% \,.\end{equation}
and we again impose the condition ${\lambda(\tilde{\lambda}=0)=0}$ to fix the
parameter ${d=\frac{c}{\alpha\beta}}$. Thus
%\begin{equation}
$\tilde{\lambda}=\frac{\alpha^2\lambda}{c-\alpha\beta\lambda}$,
% \,,\end{equation}
which finally implies
\begin{equation}
\gamma({\lambda}) = \left(
\begin{array}{c}
\alpha^2\lambda/(c-\alpha\beta\lambda)\\
\V_0\\
x_0\\
y_0\\
\end{array}\right)\,. \label{4DGeodAfBetaNeq0}
\end{equation}
\end{enumerate}
\medskip

\subsection{Background null geodesics in the $5$-dim form (\ref{5D_imp}),
  (\ref{Constraint_Hyp})} Next we connect the
 family of null geodesics (\ref{4DGeodAfBeta0}) and
 (\ref{4DGeodAfBetaNeq0}) to those of the $5$-dim parametrisation using the
 coordinates $(U,V,Z_2,Z_3,Z_4)$  of the
(A)dS universe. In general, they can be written as \cite[eq.\
(29)]{PO:01}
\begin{equation}\label{5DGeneralNullGeod}
\gamma(\lambda) = \left(
\begin{array}{c}
U^0+\dot{U}^0\,\lambda\\
V^0+\dot{V}^0\,\lambda\\
Z_2^0+\dot{Z}^0_2\,\lambda\\
Z_3^0+\dot{Z}^0_3\,\lambda\\
Z_4^0+\dot{Z}^0_4\,\lambda
\end{array}\right)\,.
\end{equation}
The integration constants are constrained by condition
(\ref{Constraint_Hyp}), its derivative with respect to the parameter
${\lambda}$, and the normalization of the $5$-velocity, that is explicitly by
\begin{align}
(Z_{2}^0)^{2}+(Z_{3}^0)^{2}+ \sigma (Z_{4}^0)^{2}-2U^0V^0&=\sigma
a^{2}\label{eq:constr-hyp}\,, \\
Z_{2}^0 \dot Z_{2}^0+Z_{3}^0 \dot Z_{3}^0+ \sigma Z_{4}^0\dot  Z_{4}^0-V^0\dot
U^0-U^0\dot V^0&=0\,,\label{eq:constr-der} \\
(\dot Z_2^0)^2+(\dot Z_3^0)^2+\sigma(\dot
Z_4^0)^2-2 \dot U^0\dot V^0&=0 \,.%e\in\{0,\pm 1\}\,,
\label{eq:norm}
\end{align}
%with $e$ being zero in the null case.%, namely ${e=0}$.

Using the simple transformation (\ref{CoordTrans_4D_to_5D}), the $5$-dim
background null geodesics corresponding to (\ref{4DGeodAfBeta0}) and
(\ref{4DGeodAfBetaNeq0})  are
\begin{equation}
  \gamma(\lambda) = \frac{1}{\alpha}\,\left(
\begin{array}{c}
\alpha^2\lambda/c	\\
\V_0\\
x_0\\
y_0\\
a(2-\alpha)
\end{array}\right),
\quad \hbox{and} \quad
\gamma(\lambda) = \frac{c-\alpha\beta\lambda}{\alpha c}\left(
\begin{array}{c}
  \alpha^2\lambda/(c-\alpha\beta\lambda)	\\
\V_0\\
x_0\\
y_0\\
a\left(2-\alpha\frac{ c}{c-\alpha\beta\lambda}\right)
\end{array}\right),
\end{equation}
respectively. We immediately observe the following:
\begin{enumerate}[noitemsep]
  \item[(1)] The constant $\beta$ in the
  second expression can be set to zero, ${\beta=0}$, to formally
obtain the first one. We can thus continue only
with the more general second expression of the $5$-dim  geodesics.
\item[(2)] The initial value of the coordinate $U$ is zero, i.e., ${U^0=0}$.
\item[(3)] W.l.o.g.\ we may set ${\dot{U}^0=1}$ corresponding to ${c=\alpha}$,
and we will do so in the following.
\end{enumerate}\medskip

Consequently, we obtain the following explicit $5$-dim  representation of the
null generators
\begin{equation}\label{5DGeodFin}
\gamma(\lambda) = \frac{1-\beta\lambda}{\alpha}\left(
\begin{array}{c}
  \alpha\lambda/(1-\beta\lambda)\\
\V_0\\
x_0\\
y_0\\
a\Big(2-\alpha/(1-\beta\lambda)\Big)
\end{array}\right)\,
\end{equation}
parametrised by the $4$-dim initial data ${\V_0,\,x_0,\,y_0}$.
Notice from the first line that, interestingly,
\begin{equation}\label{eq:U0}
  U=\lambda\,.
\end{equation}
Finally, by comparing (\ref{5DGeneralNullGeod}) with (\ref{5DGeodFin}) we can
now identify the complete set of $5$-dim data in terms of the $4$-dim
data via
\begin{align}%\label{GenPos}
&U^0=0 \,,& &V^0=\frac{\V_0}{\alpha} \,,& &Z_2^0=\frac{x_0}{\alpha} \,,&
&Z_3^0=\frac{y_0}{\alpha} \,,& &Z_4^0=a\left(\frac{2}{\alpha}-1\right) \,,
\nonumber \\
&\dot{U}^0 = 1 \,,& &\dot{V}^0=-\frac{\beta}{\alpha}\V_0 \,,&
&\dot{Z}^0_2=-\frac{\beta}{\alpha}x_0 \,,& &\dot{Z}^0_3=-\frac{\beta}{\alpha}y_0
\,,& &\dot{Z}^0_4=-2a\frac{\beta}{\alpha} \,, \label{GenVel}
\end{align}
with $\alpha$, $\beta$ given by (\ref{AlphaBetaDef}).
% , i.e.,
% \begin{equation}
% \alpha\equiv 1+\frac{\Lambda}{12}\left((x_0)^2+(y_0)^2\right) \,, \qquad\qquad \beta\equiv -\frac{\Lambda}{6}\V_0 \,. \nonumber
% \end{equation}
\medskip

\subsection{Global null geodesics in the $5$-dim  impulsive
  metric (\ref{5D_imp}), (\ref{Constraint_Hyp})}

The global geodesics in the $5$-dim  distributional representation of
non-expanding impulsive gravitational waves were first derived in \cite{PO:01}
using a formal approach. These results have been recently confirmed by a careful
regularisation analysis in the following sense \cite{SSLP:16}: The geodesics in
a \emph{general regularisation} of \eqref{5D_imp}, \eqref{Constraint_Hyp},
where the Dirac-delta is replaced by a {generic mollifier}, converge to the
distributional geodesics of \cite{PO:01}. For our purpose it suffices to deal
with the global null geodesics which have ${U^0=0}$ and ${\dot{U}^0=1}$. In the
coordinates $(U,V,Z_p)$ where $p=2,3,4$, they take the explicit form \cite[Sec.\
5]{SSLP:16}
\begin{equation}\label{5DNullGeod}
\gamma_{5D}(\lambda) = \left(
\begin{array}{c}
\lambda\\
V^0+\dot{V}^0\lambda+\Theta(\lambda)B+C\lp\\
Z_p^0+\dot{Z}^0_p\lambda+A_p\lp
\end{array}\right)\,,
\end{equation}
with $\lambda_+\equiv\lambda\Theta(\lambda)$, and ($j=2,3$)
\begin{align}\label{5DCoeff}
A_j =&\, \frac{1}{2}\Big(H_{,j}^\im+\frac{Z_j^0}{\sigma
  a^2}\left(H^\im-H_{,p}^\im Z_p^0\right)\Big), \quad
\nonumber \\
A_4 =&\, \frac{1}{2}\Big(\sigma H_{,4}^\im+\frac{Z_4^0}{\sigma
  a^2}\left(H^\im-H_{,p}^\im Z_p^0\right)\Big), \quad  \nonumber \\
B =&\, \frac{1}{2}H^\im \,,  \\\nonumber
C =&\,
\frac{1}{8}\Big((H_{,2}^\im)^2+(H_{,3}^\im)^2+\sigma(H_{,4}^\im)^2+\frac{1}{
  \sigma a^2}\big({H^\im}^2-(H_{,p}^\im Z_p^0)^2\big)\Big)
\\\nonumber
&\,+\frac{1}{2\sigma
  a^2}\left(H^\im-H_{,p}^\im Z_p^0\right)V^0+\frac{1}{2}H_{,p}^\im\dot{Z}_p^0
\,.
\end{align}
Here $H^\im$ and $H^\im,_p$ denote the values of $H$ and its respective spatial
derivatives at the instant of interaction of the geodesic with the wave
impulse, i.e., at the parameter value $\lambda=0=U$, see \eqref{eq:U0}. Also, in
the term ${H_{,p}^\im Z_p^0}$ summation over ${p=2,3,4}$ is understood.

In this general family of global null geodesics, we can identify those which
coincide with the special family of background null geodesics
\eqref{5DGeodFin} in front of the impulse by simply substituting
the constants (\ref{GenVel}) into \eqref{5DCoeff}.
\medskip

\subsection{Global null geodesics in the $4$-dim  impulsive
metric (\ref{4D_imp})}
Finally, we rewrite these global null geodesics (with the particular choice of
initial data (\ref{4DGeodAfBeta0}), (\ref{4DGeodAfBetaNeq0})) in the $4$-dim
metric form (\ref{4D_imp}). To this end, we transform (\ref{5DNullGeod}) with
\eqref{GenVel} substituted into (\ref{5DCoeff}) using the inverse of
(\ref{CoordTrans_4D_to_5D}),
\begin{equation}
\U=\Omega U \,, \qquad   \V=\Omega V \,, \qquad x=\Omega Z_2 \,, \qquad y=\Omega
Z_3 \,, \qquad \hbox{with} \qquad \Omega=\frac{2a}{Z_4+a} \,.
\label{CoordTrans_5D_to_4D}
\end{equation}
Moreover, we have to rewrite the derivatives of the function $H$ on
the impulse using $\h$ and its derivatives, see (\ref{HhRelation}),
namely\footnote{Here we use the relations ${\Lambda=3\sigma/a^2}$ and
${(x_0)^2+(y_0)^2=4\sigma
a^2(\alpha-1)}$.}
\begin{align}
H^\im_{,j}=2\h_{,j}^\im-\h^\im\,\frac{x_0^{j}}{\sigma\alpha a^2} \,, \qquad
H^\im_{,4}=-\frac{1}{a}\left(x_0\h^\im_{,x}+y_0\h^\im_{,y}\right)+2\h^\im\,\frac
{\alpha-1}{\alpha a}\,, \label{DerivativesHh}
\end{align}
where ${x^2_0\equiv x_0,\, x^3_0\equiv y_0}$. A lengthy calculation leads to
%Substituting
%(\ref{DerivativesHh}) into the coefficients (\ref{5DCoeff}) together with our
%choice of null geodesics in front of the impulse given by initial data
% (\ref{GenVel}) we obtain
\begin{align}
& A_j=\h^\im_{,j}+\frac{x_0^j}{2\sigma\alpha a^2}\, \G, \quad % \nonumber \\&
A_4=\frac{1}{\sigma\alpha a} \G, \quad %\nonumber \\&
B = \frac{1}{\alpha}\h^\im \,,\nonumber \\&
C
=\frac{1}{2}\big((\h^\im_{,x})^2+(\h^\im_{,y})^2\big)+\frac{1}{
2\sigma\alpha a^2}\big((\h^\im+\G)\V_0+\h^\im\G\big) \,, \label{5DCoeffSpec}
\end{align}
where
\begin{equation}
\G \equiv \h^\im-x_0\h^\im_{,x}-y_0\h^\im_{,y} \,.
\end{equation}
In addition, using \eqref{GenVel}, \eqref{5DNullGeod}, and
\eqref{5DCoeffSpec}, the conformal factor in \eqref{CoordTrans_5D_to_4D} takes
the form

\begin{equation}
\Omega=%\frac{2a}{Z_4+a}=
\frac{\alpha}{1-\beta\lambda+\frac{\Lambda}{6}\G\,\lp}\,.
\end{equation}
Finally, by combining (\ref{5DNullGeod}) with (\ref{5DCoeffSpec}) and
(\ref{CoordTrans_5D_to_4D}), the specific class of
global null geodesics corresponding to the solutions (\ref{4DGeodAfBeta0}),
(\ref{4DGeodAfBetaNeq0}) in front of the impulse, take the following form in
the $4$-dim  `distributional' coordinates
$(\U,\V,x,y)$ %\todo{check!!!}
\begin{align}
\!\!\!\!\gamma_{4D}[\V_0,x_0,y_0](\lambda)=\left(
\begin{array}{c}  \frac{\alpha\lambda}{1-\beta\lambda+\frac{\Lambda}{6}\G\,\lp}
\vspace{2.0mm} \\ \!\!\!\!
\V_0+\frac{\Theta(\lambda)}{1-\beta\lambda+\frac{\Lambda}{6}\G\,\lp}\,
\h^\im+\frac
{\alpha\lp}{2(1-\beta\lambda+\frac{\Lambda}{6}\G\,\lp)}\,{\mathcal F}\!\!\!\!
\vspace{2.0mm}  \\
x_0^j+\frac{\alpha\lp}{1-\beta\lambda+\frac{\Lambda}{6}\G\,\lp}\,\h_{,j}^\im\\
\end{array}\right). \label{Global4DNullGeod} \end{align}
Here, we have denoted the dependence on the initial data $[\V_0,x_0,y_0]$
explicitly, and we have used the abbreviation
\begin{equation}
  {\mathcal F}\equiv(\h_{,x}^\im)^2+(\h_
  {,y}^\im)^2+\frac{\h^\im}{\sigma\alpha a^2}(\V_0+\G)\,.
\end{equation}
Recall also that the constants $\alpha$, $\beta$ are given by
(\ref{AlphaBetaDef}).
%,
%\begin{equation}
%\alpha\equiv 1+\frac{\Lambda}{12}\left((x_0)^2+(y_0)^2\right) \,, \qquad\qquad
%\beta\equiv -\frac{\Lambda}{6}\V_0 \,. \nonumber
%\end{equation}
\medskip

At this point, we infer from the first line  in \eqref{Global4DNullGeod}  that
\begin{align}\label{eq:U}
\U &= \frac{\alpha\lambda}{1-\beta\lambda+\frac{\Lambda}{6}\G\,\lp} \,.
\end{align}
Observe that (since ${\U=\Omega U=\Omega \lambda}$, and ${\Omega>0}$) $\U$ and
$\lambda$ have the \emph{same sign}, and vanish simultaneously on the impulse.
This implies $\Theta(\U)=\Theta(\lambda)$, and together with \eqref{eq:U}
we obtain
\begin{equation}
  \Up =
  \frac{\alpha\lp}{1-\beta\lambda+\frac{\Lambda}{6}\G\,\lp}%\quad
%   \mbox{and}\quad
%   \Theta(\U)=\frac{1-\beta\lambda+\frac{\Lambda}{6}\G\,\lambda}{
%    1-\beta\lambda+\frac{\Lambda}{6}\G\,\lp}\,
%  \Theta(\lambda) 
  \,.
\end{equation}
This allows us to finally express our special family of
null geodesics (\ref{Global4DNullGeod}) in the simple explicit form
\begin{align}
\!\!\!\!\gamma_{4D}[\V_0,x_0,y_0](\U)=\left(
\begin{array}{c}
\U \vspace{2.0mm} \\
\V_0+\Theta(\U)\,\h^\im+\Up
\,\frac{1}{2}\big((\h_{,x}^\im)^2+(\h_{,y}^\im)^2\big) \vspace{2.0mm}  \\
x_0^j+\Up\h_{,j}^\im\\
\end{array}\right)\!. \label{Global4DNullGeod_uParam}
\end{align}
In the following section, we will use this final result to clarify the
geometric nature of the comoving `continuous' coordinate system and the
related Penrose junction conditions for the entire class of non-expanding
impulsive gravitational waves in (A)dS backgrounds.

\section{Cut-and-paste with $\Lambda$}\label{sec:sap}
Let us now discuss the results derived in the above section. Most importantly, 
the final formula \eqref{Global4DNullGeod_uParam} tells us that the 
discontinuous transformation \eqref{trans} is closely related to special 
geodesics in the distributional spacetime \eqref{4D_imp}, namely the null 
geodesic generators of the (A)dS hyperboloid. Indeed, the explicit form 
\eqref{Global4DNullGeod_uParam} of these geodesics, which are jumping in the 
$\V$-coordinate, and suffer a kink in  $\V$ as well as in the transverse spatial 
coordinates ${(x,y)}$, \emph{matches exactly} the transformation \eqref{trans}.

To be more precise, we can employ \eqref{Global4DNullGeod_uParam}
to transform the coordinates $(u,v,Z)\equiv(u,v,X,Y)$ in which the metric is
\emph{continuous} (cf.\ \eqref{conti}) to the coordinates
$(\U,\V,\eta)\equiv(\U,\V,x,y)$ in which the metric is \emph{distributional}
(cf.\ \eqref{4D_imp}) via
\begin{equation}\label{eq:final}
  \left(\begin{array}{c}
    u\\v\\X\\Y
   \end{array}\right)\,
   \mapsto\, \gamma_{4D}[v,X,Y](u)=
   \left(\begin{array}{c}
     u\\v+\Theta(u)\,\h^\im+\up
     \,\frac{1}{2}\big((\h_{,X}^\im)^2+(\h_{,Y}^\im)^2\big) \\
     X+\up\h_{,X}^\im\\ Y+\up\h_{,Y}^\im
   \end{array}\right)
   =
   \left(\begin{array}{c}
     \U\\\V\\x\\y
   \end{array}\right)\,,
\end{equation}
with the standard relations $Z=\frac{1}{\sqrt{2}}(X+iY)$
and $\eta=\frac{1}{\sqrt{2}}\,(x+iy)$. Observe also from \eqref{eq:H} that
\begin{equation}
  \h^\im=\h(\gamma_{4D}[v,X,Y](0))=\h(X,Y)=\h(x,y)|_{u=0}=h(X,Y)\equiv h(Z,\bar
Z)\,,
\end{equation}
and similarly for the spatial derivatives. Hence \eqref{eq:final} fully
coincides with \eqref{trans}, and we have thus \emph{derived} the `discontinuous
transformation' from a special family of unique global null geodesics of the
impulsive wave in the distributional form, obtained previously by a general
regularisation procedure in \cite{SSLP:16}.

Moreover, we explicitly see that the transformation \eqref{trans} respectively
\eqref{eq:final} \emph{turns the special geodesics
\eqref{Global4DNullGeod_uParam} into coordinate lines}, hence the coordinates
$(u,v,X,Y)$ are \emph{comoving} with the corresponding null particles. In fact,
that is the way how the distributional metric \eqref{4D_imp} in coordinates
$(\U,\V,x,y)$ is transformed to the much more regular, (locally Lipschitz)
continuous metric \eqref{conti} in the coordinates $(u,v,X,Y)$. This generalises
(and is in perfect agreement with) the flat case ${\Lambda=0}$, in which the
Rosen coordinates for \emph{pp}-waves are again comoving, cf.\ \cite{Ste:98,KS:99a}.
\medskip

\begin{figure}[htb]\centering
 \includegraphics[width=.65\textwidth]{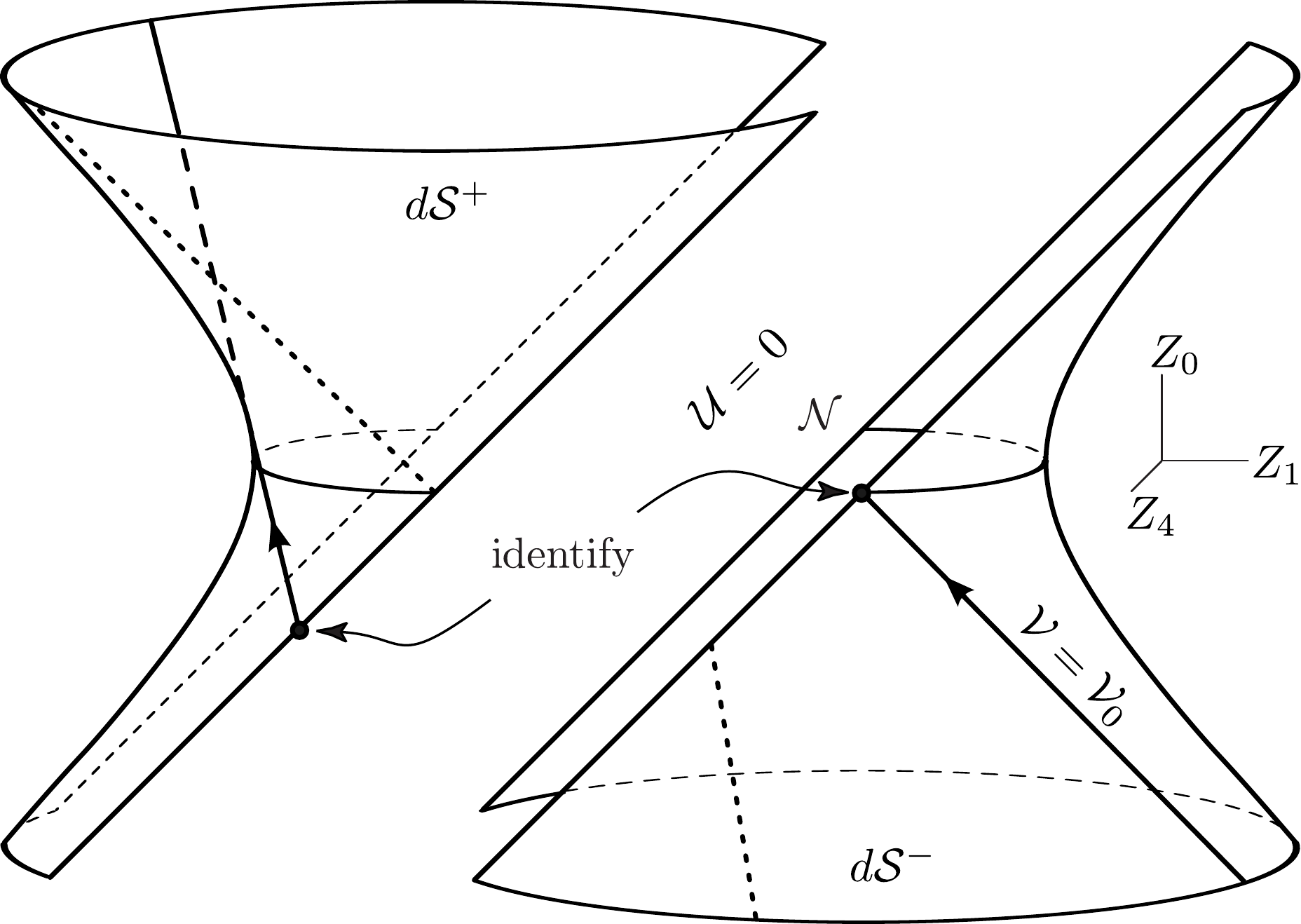}
 \caption{The construction of non-expanding spherical impulsive waves with the
cut-and-paste method: de Sitter space is cut along the null hyperplane
$\mathcal{N}$ given by ${\U=0}$, and the two `halves' $d\mathcal{S}^\pm$ are
then re-attached with a shift along the null generators of
$\mathcal{N}$. We have also depicted the null geodesic generator of
$d\mathcal{S}^-$ given by ${\V=\V_0}$, hitting $\mathcal{N}$. Instead of continuing as
unbroken null generator into $d\mathcal{S}^+$ (indicated by the dashed line),
the interaction with the impulse does not only make it jump due to the
identification of points but also refracts it to become the appropriate null
generator of $d\mathcal{S}^+$.
}
\label{hypcut}
\end{figure}

This insight can now be put to use to arrive at a vivid picture illustrating
the cut-and-paste approach in the $\Lambda\not=0$-case, see Figure
\ref{hypcut}. Here we exclusively concentrate on the de Sitter case, the
anti-de Sitter case being analogous.

The de Sitter hyperboloid is cut into two `halves' $d\mathcal{S}^-$ and $d\mathcal
S^+$ along the non-expanding spherical impulsive surface ${U=0}$ (for its geometry see
\cite[Fig.\ 3]{PG:97}). Then the two parts are re-attached in the following
specific way: The null geodesic generators of the two halves are joined
according to \eqref{Global4DNullGeod_uParam}, or \eqref{5DNullGeod} with
(\ref{GenVel}) substituted into \eqref{5DCoeff} in the $5$-dim picture, depending on $H^\im(Z_2,Z_3,Z_4)$.
We will refer to this $5$-dim picture in the remainder of this discussion, since it is more
directly related to Figure \ref{hypcut}. More precisely, the generators
\eqref{5DGeodFin} approaching the impulse at ${\U=0}$ from $d{\mathcal S}^-$ are
shifted from the value ${\V_0/\alpha=V^0}$ in $d{\mathcal S}^-$ to the value
${V^0+\frac{1}{2}H^\im}$ in $d{\mathcal S}^+$ due to the interaction with the 
wave. They are also tilted (deflected) by the impulse according to $C$, cf.\ 
\eqref{5DNullGeod}, which is precisely the amount needed to turn them into the 
generators of ${d\mathcal S}^+$ starting at ${V^0+\frac{1}{2}H^\im}$.

Note, however, the following subtlety: In Figure \ref{hypcut} the
$Z_2$- and $Z_3$-directions are suppressed, so that the two families of null
generators plotted in each `half' (i.e., those whose only spatial movement is in
the $Z_4$-direction) \emph{need not} in general be the ones that get matched.
The reason is that the impulse induces a motion in the $Z_2$- and
$Z_3$-directions as is evident from the $\lambda_+$-term in the spatial part of
\eqref{5DNullGeod} and the nontrivial constants $A_j$ in
\eqref{5DCoeff}\footnote{Observe that this effect already occurs in
the flat case, see Figure \ref{fig:planecut}.}. However, if we specialise to
waves with ${H^\im_{,j}=0}$, the depicted generators which satisfy ${x_0=0=y_0}$
implying ${Z^0_j=0=\dot Z^0_j}$, are precisely the ones shown in 
Figure \ref{hypcut}. Indeed, in this case we have $A_j=0$ (cf.\ \eqref{GenVel}, 
\eqref{5DCoeff}) and the interaction with such an impulse does not induce any 
motion in the $Z_j$-directions since by \eqref{5DNullGeod} $Z_2(\lambda)=0= 
Z_3(\lambda)$ for all $\lambda$. So the null generators of $d{\mathcal 
S}^-$  and $d{\mathcal S}^+$ depicted in Figure \ref{hypcut} are indeed matched 
to one another.
\medskip

This explicit visualization thus provides us with a clear geometric insight. It
yields a deeper understanding of the various construction methods of impulsive
gravitational waves in de Sitter and anti-de Sitter universes. Moreover, it
gives a natural explanation on their mutual unambiguous relations, because the
key elements of the overall geometric picture --- the null generators --- are
globally unique.

\section{Discussion and outlook}

We have given a vivid interpretation of the Penrose junction conditions and the
corresponding `discontinuous coordinate transformation' which relates the
continuous and distributional metric forms of the entire class of non-expanding
impulsive gravitational waves travelling in an (A)dS background. It is based on
an explicit study of the null geodesics in the distributional $5$-dim form of
the metric, which have been  derived via a general regularisation procedure in
\cite{SSLP:16}: The (A)dS hyperboloid is cut into two `halves' along the null
wave surface ${\{U=0\}}$, which are then re-attached in such a way that the null
geodesic generators of both `halves' are matched with a shift in the
$V$-direction. The `discontinuous coordinate transformation' which turns these
broken and refracted null generators into coordinate lines also transforms the
distributional form of the metric into the continuous one.

In the case $\Lambda=0$ this notorious transformation has been treated in a
mathematically rigorous way. In \cite{KS:99a} it was shown to be connected to a
`generalised diffeomorphism' in the framework of nonlinear distributional
geometry \cite[Ch.\ 3]{GKOS:01}. The geometric insight provided by the present
work paves the way to a generalisation of this approach to the case of
nonvanishing $\Lambda$. Indeed, building on the fully nonlinear distributional
analysis of the geodesics in the $5$-dim picture \cite{SS:17} the transformation
\eqref{eq:final} can now be shown to be a (distributional) limit of a
`generalised diffeomorphism', which is built up from a regularised version of
the geodesics \eqref{5DNullGeod}. The details of this account will be studied in
a separate, more mathematically oriented paper \cite{SSS:19}.
\medskip

Finally, our current findings have some consequences for studying the wave 
memory effect in \emph{impulsive} spacetimes. In particular, the geometric 
picture established above makes it obvious that the arguments recently put 
forward in \cite{S:19} simply extend to (A)dS backgrounds: The interaction of 
test particles with the wave impulse can be seen most explicitly and the 
difference of their positions and velocities before and after the wave has 
passed can simply be read off from the \emph{matching conditions} already 
derived in \cite[Sec.\ 4]{PSSS:15}. (In particular, there is no need to involve 
the symmetries of the (A)dS background.) A comprehensive study of the memory 
effect in impulsive wave spacetimes is, however, subject to current research and 
will be detailed elsewhere.

\section*{Acknowledgement}
This work was supported by project P28770 of the Austrian Science Fund FWF, the
WTZ-grant CZ12/2018 of OeAD/the Czech-Austrian MOBILITY grant 8J18AT02, and the
Czech Science Foundation Grant No.\ GA\v CR 19-01850S.

\bibliographystyle{abbrv}
\bibliography{ro}

\begin{thebibliography}{10}

\bibitem{AS:71}
P.~C. Aichelburg and R.~U. Sexl.
\newblock On the gravitational field of a massless particle.
\newblock {\em Gen.\ Rel.\ Grav.}, { 2}:303--312, 1971.

\bibitem{BH:03}
C.~Barrab\`es and P.~A. Hogan.
\newblock {\em Singular null hypersurfaces in general relativity}.
\newblock World Scientific Publishing Co., Inc., River Edge, NJ, 2003.

\bibitem{CG:12}
P.~T. Chru{\'s}ciel and J.~D.~E. Grant.
\newblock On {L}orentzian causality with continuous metrics.
\newblock {\em Class.\ Quant.\ Grav.}, 29(14):145001, 32, 2012.

\bibitem{GT:87}
R.~Geroch and J.~Traschen.
\newblock Strings and other distributional sources in general relativity.
\newblock {\em Phys.~Rev.~D}, {36}(4):1017--1031, 1987.

\bibitem{GGKS:18}
M.~Graf, J.~D.~E. Grant, M.~Kunzinger, and R.~Steinbauer.
\newblock The {H}awking-{P}enrose singularity theorem for
  {$C^{1,1}$}-{L}orentzian metrics.
\newblock {\em Comm. Math. Phys.}, 360(3):1009--1042, 2018.

\bibitem{GKSS:19}
J.~D.~E. Grant, M.~Kunzinger, C.~S\"amann, and R.~Steinbauer.
\newblock The future is not always open.
\newblock {\em preprint, arXiv:1901.07996 [math.DG]}, 2019.

\bibitem{GP:09}
J.~B. Griffiths and J.~Podolsk\'y.
\newblock {\em {Exact Space-Times in Einstein's General Relativity}}.
\newblock Cambridge University Press, Cambridge, 2009.

\bibitem{GKOS:01}
M.~Grosser, M.~Kunzinger, M.~Oberguggenberger, and R.~Steinbauer.
\newblock {\em Geometric theory of generalized functions with applications to
  general relativity}, volume 537 of {\em Mathematics and its Applications}.
\newblock Kluwer Academic Publishers, Dordrecht, 2001.

\bibitem{HW:51}
P.~Hartman and A.~Wintner.
\newblock On the problems of geodesics in the small.
\newblock {\em Amer. J. Math.}, 73:132--148, 1951.

\bibitem{HT:93}
M.~Hotta and M.~Tanaka.
\newblock Shock wave geometry with nonvanishing cosmological constant.
\newblock {\em Class.\ Quant.\ Grav.}, {10}:307--314, 1993.

\bibitem{KS:99}
M.~Kunzinger and R.~Steinbauer.
\newblock {A note on the Penrose junction conditions}.
\newblock {\em Class. Quant. Grav.}, 16:1255--1264, 1999.

\bibitem{KS:99a}
M.~Kunzinger and R.~Steinbauer.
\newblock A rigorous solution concept for geodesic and geodesic deviation
  equations in impulsive gravitational waves.
\newblock {\em J. Math. Phys.}, 40(3):1479--1489, 1999.

\bibitem{KSS:14}
M.~Kunzinger, R.~Steinbauer, and M.~Stojkovi{\'c}.
\newblock The exponential map of a {$C^{1,1}$}-metric.
\newblock {\em Differential Geom. Appl.}, 34:14--24, 2014.

\bibitem{KSSV:14}
M.~Kunzinger, R.~Steinbauer, M.~Stojkovi{\'c}, and J.~A. Vickers.
\newblock A regularisation approach to causality theory for
  {$C^{1,1}$}-{L}orentzian metrics.
\newblock {\em Gen.\ Rel.\ Grav.}, 46(8):1738, 18, 2014.

\bibitem{LS:94}
C.~O. Lousto and N.~G. Sanchez.
\newblock {Scattering processes at the Planck scale}.
\newblock In {\em {2nd Journee Cosmologique within the framework of the
  International School of Astrophysics, D. Chalonge Paris, France, June 2-4,
  1994}}, pages 339--370, 1994.

\bibitem{Min:15}
E.~Minguzzi.
\newblock Convex neighborhoods for {L}ipschitz connections and sprays.
\newblock {\em Monatsh. Math.}, 177(4):569--625, 2015.

\bibitem{Min:19}
E.~Minguzzi.
\newblock Causality theory for closed cone structures with applications.
\newblock {\em Rev.\ Math.\ Phys.}, 31(5):930001, 139, 2019.

\bibitem{P:68a}
R.~Penrose.
\newblock Structure of space-time.
\newblock In {\em Battelle Rencontres, 1967 Lectures in Mathematics and
  Physics}, pages 121--235. Benjamin, New York, 1968.

\bibitem{P:68}
R.~Penrose.
\newblock {Twistor quantization and curved space-time}.
\newblock {\em Int. J. Theor. Phys.}, 1:61--99, 1968.

\bibitem{Pen:72}
R.~Penrose.
\newblock The geometry of impulsive gravitational waves.
\newblock In {\em General relativity (papers in honour of {J}. {L}. {S}ynge)},
  pages 101--115. Clarendon Press, Oxford, 1972.

\bibitem{P:98}
J.~Podolsk\'{y}.
\newblock Non-expanding impulsive gravitational waves.
\newblock {\em Class.\ Quant.\ Grav.}, 15(10):3229--3239, 1998.

\bibitem{P:02}
J.~Podolsk\'y.
\newblock {Exact impulsive gravitational waves in space-times of constant
  curvature}.
\newblock In {\em Gravitation: Following the Prague Inspiration}, pages
  205--246. Singapore: World Scientific Publishing Co., 2002.

\bibitem{PG:97}
J.~Podolsk\'{y} and J.~B. Griffiths.
\newblock Impulsive gravitational waves generated by null particles in de
  {S}itter and anti-de {S}itter backgrounds.
\newblock {\em Phys. Rev. D (3)}, 56(8):4756--4767, 1997.

\bibitem{PG:99a}
J.~Podolsk\'{y} and J.~B. Griffiths.
\newblock Impulsive waves in de {S}itter and anti-de {S}itter spacetimes
  generated by null particles with an arbitrary multipole structure.
\newblock {\em Class.\ Quant.\ Grav.}, 15(2):453--463, 1998.

\bibitem{PG:99}
J.~Podolsk\'{y} and J.~B. Griffiths.
\newblock Nonexpanding impulsive gravitational waves with an arbitrary
  cosmological constant.
\newblock {\em Phys. Lett. A}, 261(1-2):1--4, 1999.

\bibitem{PO:01}
J.~Podolsk\'{y} and M.~Ortaggio.
\newblock Symmetries and geodesics in (anti-) de {S}itter spacetimes with
  non-expanding impulsive waves.
\newblock {\em Class.\ Quant.\ Grav.}, 18(14):2689--2706, 2001.

\bibitem{PSSS:15}
J.~Podolsk{\'y}, C.~S{\"a}mann, R.~Steinbauer, and R.~{\v{S}}varc.
\newblock The global existence, uniqueness and {${C}^1$}-regularity of
  geodesics in nonexpanding impulsive gravitational waves.
\newblock {\em Class.\ Quant.\ Grav.}, 32(2):025003, 23, 2015.

\bibitem{PV:98}
J.~Podolsk\'y and K.~Vesel\'y.
\newblock Continuous coordinates for all impulsive pp-waves.
\newblock {\em Phys.\ Lett.\ A}, {241}:145--147, 1998.

\bibitem{SS:17}
C.~S\"amann and R.~Steinbauer.
\newblock Geodesics in nonexpanding impulsive gravitational waves with
  {$\Lambda$}. {II}.
\newblock {\em J. Math. Phys.}, 58(11):112503, 18, 2017.

\bibitem{SS:18}
C.~S\"amann and R.~Steinbauer.
\newblock On geodesics in low regularity.
\newblock {\em Journal of Physics: Conference Series}, 968:012010, 14, 2018.

\bibitem{SSLP:16}
C.~S\"amann, R.~Steinbauer, A.~Lecke, and J.~Podolsk\'y.
\newblock Geodesics in nonexpanding impulsive gravitational waves with
  {$\Lambda$}, part {I}.
\newblock {\em Class.\ Quant.\ Grav.}, 33(11):115002, 33, 2016.

\bibitem{SSS:19}
C.~S\"amann, R.~Steinbauer, and R.~\v{S}varc.
\newblock Cut-and-paste for impulsive gravitational waves with {$\Lambda$}: The
  rigorous account.
\newblock {\em in preparation}, 2019.

\bibitem{S:18}
G.~M. Shore.
\newblock {Memory, Penrose limits and the geometry of gravitational shockwaves
  and gyratons}.
\newblock {\em Journal of High Energy Physics}, 2018(12):133, Dec 2018.

\bibitem{Ste:98}
R.~Steinbauer.
\newblock On the geometry of impulsive gravitational waves.
\newblock {\em ArXiv:9809054[gr-qc]}, 1998.

\bibitem{Ste:14}
R.~Steinbauer.
\newblock Every {L}ipschitz metric has {$C^1$}-geodesics.
\newblock {\em Class.\ Quant.\ Grav.}, 31(5):057001, 3, 2014.

\bibitem{S:19}
R.~Steinbauer.
\newblock Comment on `memory effect for impulsive gravitational waves'.
\newblock {\em Class.\ Quant.\ Grav.}, 36(9):098001, 2019.

\end{thebibliography}
\end{document}